\documentclass[12pt]{article}
\usepackage{graphicx}
\usepackage{color}
\usepackage{fullpage}
\usepackage{bm}% bold math
\begin{document}
%%\documentclass[referee]{jfm_oat15a}
%\documentclass[referee]{jfm}
%\usepackage[dvips]{epsfig}
%\usepackage{graphicx}
%\usepackage{epstopdf}
%\usepackage{natbib}
%              }
\title{A stochastic view of isotropic turbulence decay}
\author{M. Meldi, P. Sagaut \& D. Lucor}
\date{2011}
\maketitle
\begin{abstract}
A stochastic EDQNM approach is used to investigate  self-similar decaying isotropic turbulence at high Reynolds number ($400 \leq Re_\lambda \leq 10^4$).
The realistic energy spectrum functional form recently proposed by \cite{Meyers2008} is generalised by considering some of the model constants as random parameters, since they escape measure in most experimental set-ups. The induced uncertainty on the solution is investigated building response surfaces for decay power-law exponents of usual physical quantities.
Large-scale uncertainties are considered, the emphasis being put on Saffman and Batchelor turbulence.
The sensitivity of the solution to initial spectrum uncertainties is quantified through probability density functions of the decay exponents.
It is observed that initial spectrum shape at very large scales governs the long-time evolution, even at high Reynolds number, a parameter which is not explicitly taken into account in many theoretical works. Therefore, a universal asymptotic behavior in which kinetic energy decays as $t^{-1}$ is not detected.
But this decay law is observed at finite Reynolds number with low probability for some initial conditions. %
\end{abstract}
\noindent ~\hrulefill
\section{Introduction}
\label{introDecay}
The decay of isotropic turbulence is one of the oldest topic considered in the field of turbulence theory, which started with the seminal studies of \cite{Taylor1935}.
Comprehensive reviews are presented by \cite{Batchelor1953}, \cite{Hinze1975}, \cite{Monin75}, \cite{Davidsonbook} and \cite{Sag2008}.
One of the most famous related issue deals with the existence and uniqueness of self-similar regimes, in which global turbulent quantities, such as turbulence kinetic energy $q$, behave like
\begin{equation}
\label{eq:1}
q (x) =   A \left(  x/M_u - x_0/M_u \right) ^{n} \quad \mbox{or} \quad  q (t) =  A'  \left(  t - t_0 \right) ^{n}
\end{equation}
where $x, x_0, M_u, A, A' $ and $n$ are the distance downstream the grid, the virtual origin of isotropic turbulence, the mesh size of the grid used to trigger turbulence, two  parameters and the decay exponent, respectively. Equivalence between space- and time-dependent decay is recovered thanks to  Taylor's frozen turbulence hypothesis.\\
Influential works of \cite{Kolm1941}, \cite{Batch1956} and \cite{Saff1967} have shown that the value of the decay exponent value $n$ ( or  equivalent power law exponent for other global physical quantities) is not unique but mainly governed by initial conditions (see also \cite{Lavoie2007}, \cite{Uberoi1963}). Theory of self-similar decay has been addressed by many authors, among them \cite{george1992} and \cite{speziale1992}, in which an explicit dependency with respect to initial conditions is taken into account.
More specifically, the shape of the spectrum at very large scales is observed to be of primary importance. In fact several spectrum shapes at very large scale are known to be physically realizable in isotropic turbulence. This point has been discussed by many authors, since it is related to the famous controversy dealing with existence of invariants in high-Reynolds decaying turbulence. \cite{Saff1967} showed that the invariance of the Birkhoff-Saffman invariant $L = \int <\mathbf{u \cdot u'} > d \mathbf{r}$ during turbulence decay is related to linear momentum conservation. The associated kinetic energy spectrum behaves like ${E}(k \rightarrow 0)= Lk^2/ 4 \pi^2$ and  theoretical analysis shows that the turbulent kinetic energy decays as $q(t) \sim t^{-6/5}$. If $L=0$ the resulting condition is referred to as Batchelor turbulence: the corresponding large scales behaves like ${E}(k \rightarrow 0)= Ik^4/ 24 \pi^2$, where $I = \int \mathbf{r}^2 <\mathbf{u \cdot u'} > d \mathbf{r}$ is the Loitsyansky's integral. In this case, the turbulent kinetic energy scales as $q(t) \sim t^{-10/7}$.\\
The Reynolds number is also known to be a key parameter. In fact, low-Reynolds number flows, which are mostly driven by viscous linear effects, do not exhibit the same power-law exponent as high-Reynolds number flows (see e.g.  \cite{Batchelor1948,Burattini2006} ).

Prediction of power-law exponent starting from a simplified kinetic energy spectrum shape was introduced by \cite{Benn1966}, who considered a two-range spectrum and dimensional analysis. This method was then revised by \cite{Saff1967,Saff1967b}. Following these works, the spectrum is divided into two power-law ranges  joining at a peak located at $k_L$ : one for the very large scales, ${E}(k \leq k_L) = A k^\sigma, \sigma \in [1,4]$, and a Kolmogorov-type inertial range at smaller scales ${E}(k \geq k_L) = C_K \varepsilon ^{2/3} k^{-5/3}$, where $ \sigma, C_K$ and $\varepsilon$ denote the spectrum slope at large scales , the Kolmogorov constant and the turbulent dissipation rate, respectively. While Saffman ($\sigma = 2$) and Batchelor ($\sigma = 4$) turbulences have been extensively analysed, other possibilities have also been considered, e.g. $\sigma = 1$ and $\sigma = 3$ in \cite{Clark1998}, \cite{Ober2002}. Predictions dealing with power-law exponents of some usual physical quantities obtained using the two-range spectrum model in the high-Reynolds number case are presented  in the first column of Table \ref{tab::CBC-1}. This analysis was then further developed to investigate the sensitivity of the power-law exponent to additional features of the initial condition. \cite{Skrbrek2000} considered a smoothed form of the spectrum at $k_L$, along with internal intermittency, clipping in the dissipation range and large-scale saturation effects. An important conclusion  is that saturation induced by boundary condition yields a dramatic change in the power-law exponent, making it very difficult to distinguish between satured high-Reynolds regime and free low-Reynolds evolution. Truncation of energy spectrum at very large scales also corrupts the computation of integral quantities, such as integral length scale \cite{wang2002}.
The use of two power-law ranges was also shown to be inconsistent with the existence of an helicity-based invariant in the limit of infinitely high Reynolds number, a three-range model being necessary to this end, see \cite{Frenkel1983} and \cite{Frenkel1984}. 
%Frenkel's three-range model was later used by \cite{Eyink2000} to discuss the possible breakdown of self-similarity.\\
%
\begin{table} \centering \begin{tabular*} {\textwidth}{@{\extracolsep{\fill}} l c c c c c } \smallskip
Quantity& CBC formula &$\sigma=1$& $\sigma=2$& $\sigma=3$& $\sigma=4$ \cr \hline
$q$       &$\displaystyle -2 \frac{\sigma +1}{\sigma +3}$     &-1.005 ($<$1\%)   &-1.213 (1.2\%) &-1.342 ($<$1\%)    & -1.4014 (2\%) \cr
$L$      &$\displaystyle \frac{2}{\sigma +3}$      &0.457 (8\%) &0.402 ($<$1\%) &0.336   (1.8\%)      & 0.307 (7.8\%) \cr
$\eta$      & $\displaystyle \frac{3 \sigma +5}{4(\sigma +3)}$   &0.502  (1\%) &0.556 (1.2\%)  & 0.588    ($<$1\%)   & 0.603 ($<$1\%)  \cr
$\lambda$       &$1/2$     & 0.501 ($<$1\%)   &0.505 (1\%) & 0.505 (1\%)    & 0.505 (1\%) \cr
$\varepsilon$ & $\displaystyle \frac{-3 \sigma -5}{\sigma +3}$ &-2.009 ($<$1\%) &-2.226 (1.2\%) & -2.352 ($<$1\%) & -2.412 ($<$1\%) \cr
$Re_{\lambda}$ & $\displaystyle \frac{1 - \sigma}{2(\sigma +3)}$ &-0.0006  &-0.1006 ($<$1\%)  &-0.166 ($<$1\%) & -0.195 (8.9\%) \cr
$Re_{L}$   &  $\displaystyle \frac{1 - \sigma}{\sigma +3}$   &-0.044 &-0.204 (2.3\%) &-0.335 ($<$1\%) & -0.393 (8.3\%) \cr \hline
 \end{tabular*} \caption{Power-law exponents in high-Reynolds isotropic turbulence decay. Comte--Bellot-Corrsin (CBC) formula denotes expressions obtained via dimensional analysis. Other columns display values computed using classical deterministic EDQNM simulations, using a two-range Comte--Bellot-Corrsin spectrum at initial time and ($Re_\lambda (t=0) = 10^4$). Relative error with respect to the theoretical predictions are reported between parentheses. $q$: turbulent kinetic energy; $L$: integral lengthscale; $\eta$: Kolmogorov lengthscale; $\lambda$: Taylor micro-scale; $\varepsilon$: turbulent dissipation rate; $Re_\lambda =\sqrt{2q/3}\lambda/\nu$; $Re_L = \sqrt{2q/3}L/\nu$.}\label{tab::CBC-1} \end{table}
~\\ %
An important point is that experimental validation of theoretical predicted behaviour is elusive for many reasons. First, it is almost impossible to enforce the spectrum shape at very large scales in laboratory experiments. Second, in most cases, the large-scale spectrum shape is not directly measured but deduced from the measured decay law and theoretical relations which bridge between them. Third, identification of the three free parameters in Eq. (\ref{eq:1}), namely $A$ (or $A'$), $x_0$ and $n$, leads to the definition of a non-robust optimisation problem, which introduces a significant uncertainty in the experimental estimates as discussed by \cite{Moha1990} and  \cite{Krog2010}. Finally, it has been recently shown that spurious saturation effects may occur in numerical simulations due to the use of periodic boundary conditions, and that very large computational domains must be used in order to preclude such problem, \cite{Ish2006}. Looking at open literature, it appears that large-enough domains have not been used in most published papers, precluding definitive analysis.
Another important issue is the possible existence of a universal decay regime, in which kinetic energy should decay as $t^{-1}$. Such a regime is predicted by some theoretical models like \cite{george1992} and \cite{speziale1992}, but it has not been observed up to now and seems at least partially contradictory with a strong dependency upon initial conditions.\\
The aim of the present paper is to further investigate the sensitivity of the power-law exponents with respect to details of the initial kinetic energy spectrum. To this end, the  spectrum shape function  very recently introduced by \cite{Meyers2008} is considered.
This model, which accounts for both intermittency and bottleneck effects, includes several intrinsic parameters, which escape measurement in laboratory experiments.
Therefore, they must be considered as epistemic uncertainties and modeled as random variables.
 A direct consequence of dependency with respect to initial conditions is that a stochastic approach must be used instead of the usual deterministic one, and that power-law exponents must be characterized using numerical tools streaming from probability theory. In the present paper, the response surfaces of power-law exponents of several quantities of interest (kinetic energy $q$, integral scale $L$, Kolmogorov scale $\eta$) are constructed as a polynomial approximation obtained from a stochastic spectral projection method (\cite{Ko_Lucor_Sagaut_POF_2008}).
Samples are generated using a high-fidelity model whose accuracy has been assessed for isotropic turbulence decay, namely the EDQNM model (\cite{orszag70,Sag2008}). EDQNM solver makes it possible to account for fine details of the kinetic energy spectrum.\\
The paper is structured as follows. The turbulent energy spectrum model used in the present study is presented in Section \ref{EnergySp}, along with the response-surface building strategy, which is based on the Polynomial Chaos representation (\cite{GhanemS91}). High-Reynolds decay with uncertain spectrum shape at large scales is investigated. In Section \ref{resuHigh}, the spectrum slope parameter $\sigma$ itself is considered as a random parameter, in order to mimic experimental uncertainties. Then, the influence of other large-scale parameters   for both Saffman and Batchelor turbulence is further analyzed in Section \ref{bottleHigh}. 
\section{Energy spectrum model and response-surface parameterization}
\label{EnergySp}
The determination of the turbulent energy spectrum shape is an old but still very active research topic. Many models have been proposed (see for instance \cite{Hinze1975}, \cite{Monin75}), varying in complexity and approaches, spanning from simple definitions to more accurate models  with the drawback of introducing a significant number of free parameters. 
In the present work,  the  energy spectrum model proposed by  \cite{Meyers2008} is used, since it accounts for all known features of the kinetic energy spectrum. 
A convenient approach is to express them as functionals of random variables. It is written as follows: 
\begin{equation}
\label{eq:def-spectrum-Pope}
E(k)= C_K \varepsilon^{2/3}k^{-5/3}(kL)^{- \beta}f_L(kL)f_{\eta}(k \eta)
\label{eq::spectrum}
\end{equation}
where  $\beta$ is the intermittency correction in the inertial range of the spectrum.  It is taken equal to $\beta = \mu /9$, with $\mu = 0.25$ being the value commonly found in literature.  
Following \cite{Meyers2008}, the Kolmogorov constant $C_K$  is taken equal to 2.0173, 
$L= E_t^{3/2} / \varepsilon$ is the integral length scale and $\eta$ is the Kolmogorov length scale. The functions $f_L$ and $f_{\eta}$ shape the spectrum at very large and very small scales, respectively. They are expressed as:
\begin{equation}
f_L( k L) =  \left(\frac{kL}{[(kL)^p +\alpha_5]^{1/p}} \right)^{5/3 + \beta + \sigma}, \quad f_{\eta}( k \eta) = e^{ -\alpha_1 k \eta}  B(k \eta)
\end{equation}
The  {\it bottleneck} correction $B( k \eta)$ is 
\begin{equation}
B( k \eta) = 1+ \frac{\alpha_2(k \eta / \alpha_4)^{\alpha_3}}{1+(k \eta / \alpha_4)^{\alpha_3}}
\label{eq::bottle}
\end{equation}
The two shape functions introduce five arbitrary parameters. Parameters  $\alpha_1 - \alpha_4$ govern the shape of the spectrum at high wave numbers while $\alpha_5$ controls it at low wavenumbers. 
In \cite{Meyers2008}, it is proposed to compute these parameters by solving a system of five equations, which are related to the recovery of target values for the turbulent kinetic energy, the turbulent dissipation (or equivalently the enstrophy), the palinstrophy (or equivalently the longitudinal velocity derivative skewness) and constraints derived considering some possibly universal feature of the dissipation spectrum.\\
In Section \ref{resuHigh}, large-scale uncertainty is modeled considering $\sigma$, $p$ and $\alpha_5$  as independent random variables with  {\em uniform} distribution, the range of variation being adjusted based on open literature. The choice of uniform distributions means that we do not favor any particular parametric values within the ranges of interest. Coefficients  $\alpha_1 - \alpha_4$ are estimated using the  procedure proposed in \cite{Meyers2008} for each realization, enforcing the same global quantities (enstrophy, palinstrophy, position and value of the compensated dissipation spectrum peak) at the initial time.
Large-scale dependency of Batchelor and Saffman turbulence are further analyzed fixing $\sigma$ and considering $p$, $\alpha_5$ and $\mu$ as uncertain parameters.
The response surface of the solution with respect to uncertain random variables is built in the present work using the generalized Polynomial Chaos (gPC) method, which is a non-statistical method used to solve stochastic differential (SDE) and stochastic partial differential equations (SPDE).This method was recently applied in the field of turbulence research by \cite{lucorjfm}.
Following this approach, the power law decay exponent of each physical quantity under consideration, which is now a random variable depending nonlinearly on the uncertain parameters, can be represented thanks to the following polynomial expansion with pseudo-spectral accuracy:
%
%\begin{equation}
%E(k,t; \boldsymbol{\xi})=\sum_{l=0}^{\infty}E_l(k,t)\Phi_l(\boldsymbol{\xi})
%\end{equation}
%
\begin{equation}
n(\boldsymbol{\xi})=\sum_{l=0}^{\infty}n_l\Phi_l(\boldsymbol{\xi})
\end{equation}
where $\Phi _l$ is an orthogonal polynomial set with respect to the joint pdf of the random array $\boldsymbol{\xi}$ whose components are the independent random variables used to model the parametric uncertainty.
In practical cases, the expansion is truncated and the expansion coefficients $n_l$ are computed using a Galerkin-type projection method relying on a multidimensional Gaussian quadrature.
In the present work, all random variables were assumed to follow a uniform distribution, leading to the use of Legendre polynomials. 
It was checked, that in all cases, the use of third order polynomials along with four quadrature points (per dimension) yields fully converged results.
It is reminded that a deterministic EDQNM simulation is performed at each quadrature point, leading to $4^N$ simulations, where $N$ is the number of uncertain parameters under consideration. For each quadrature point in the uncertainty space, an energy spectrum fitting relations (\ref{eq::spectrum}) - (\ref{eq::bottle}) is set as initial condition, a self-similar regime with the same time-independent spectrum shape coefficients and decay exponent is obtained via EDQNM simulation and the corresponding power law coefficients are then measured. The gPC method is then applied as a post processing tool, reconstructing the response surface of the power law coefficients over the continuous uncertainty space starting from the discrete values computed at the quadrature points. 
The EDQNM deterministic solver was validated taking a simplified two-range CBC spectrum as initial condition for $\sigma =1, 2, 3$ and $4$ at high Reynolds number ($Re_\lambda (t=0) =10^4$). The results where very close to those recovered using simple dimensional-analysis-based calculations (see Table \ref{tab::CBC-1}). Discrepancies observed in the case $\sigma = 4$ stem from the fact that, in Batchelor turbulence, EDQNM predicts the existence of a time drift of the coefficient $A$ in $L(t) = A (t-t_0)^n$, which is not distinguished from the power-law behaviour. 
An important point is that $\lambda \sim t^{1/2}$ is recovered within 1\% error on the exponent in all cases.
\section{High Reynolds turbulence decay with uncertain large-scale spectrum shape}
\label{resuHigh}
\begin{figure}
 \centering \includegraphics[width=0.28\linewidth]{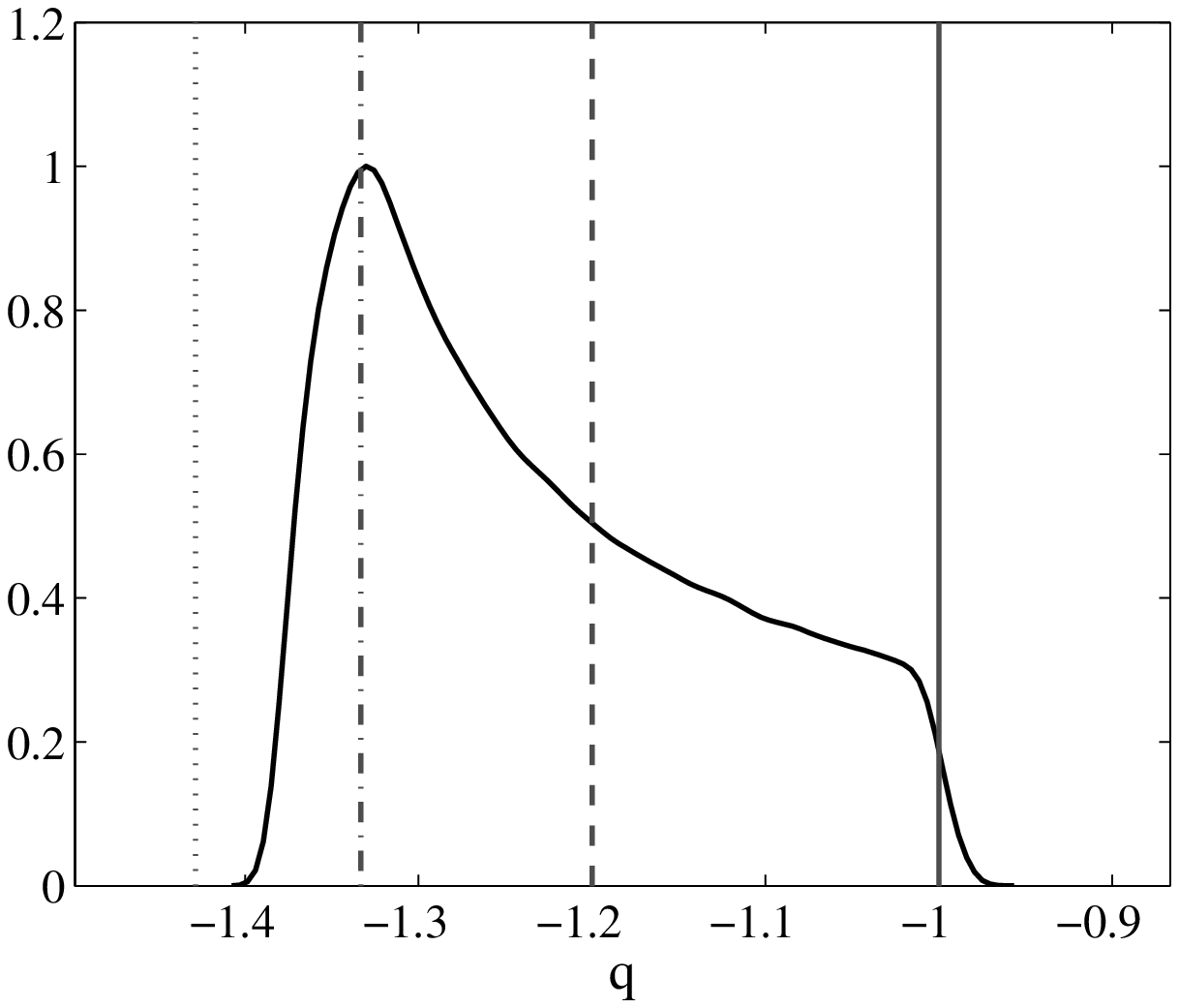} 
 						\includegraphics[width=0.28\linewidth]{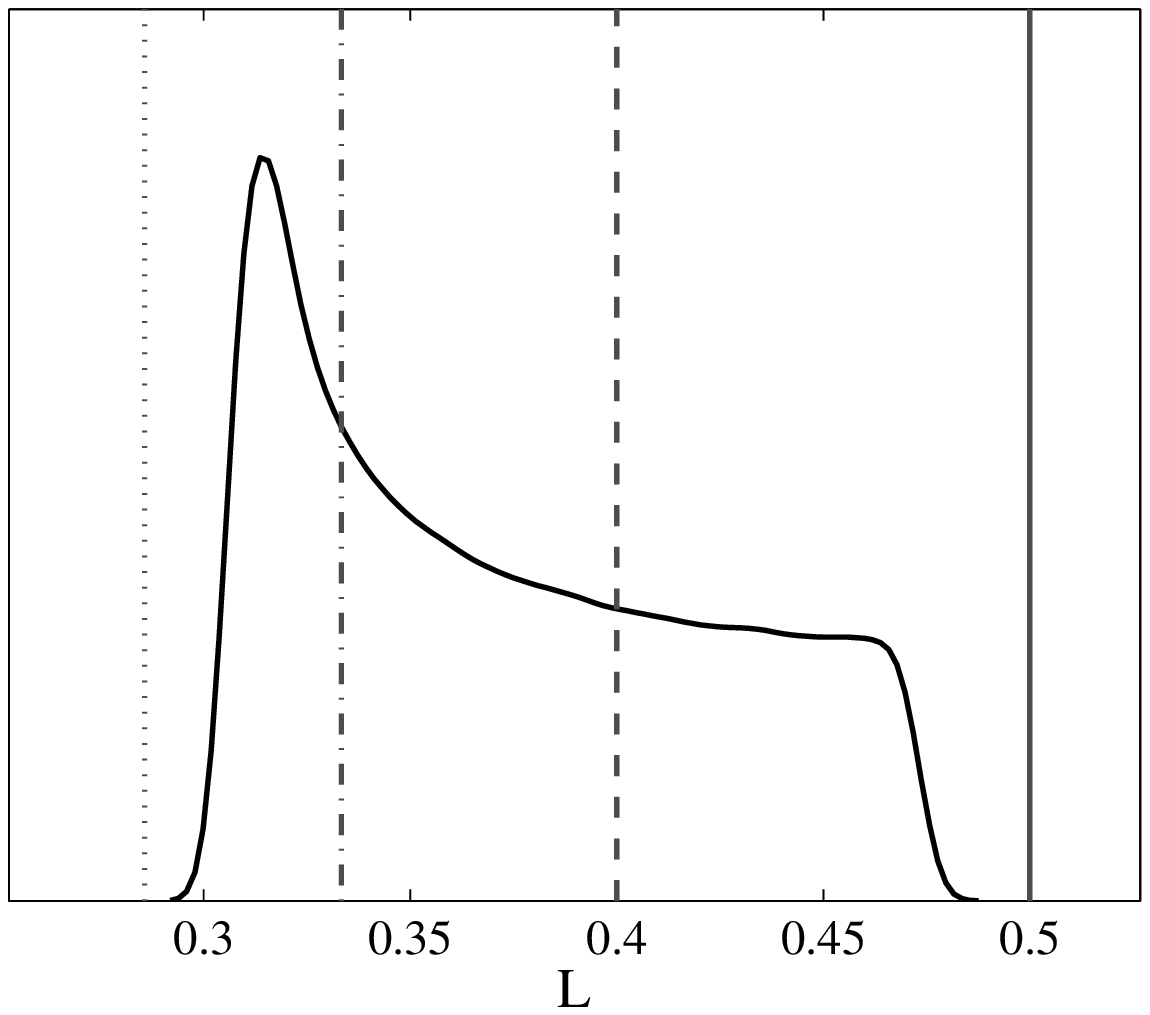}
 						\includegraphics[width=0.28\linewidth]{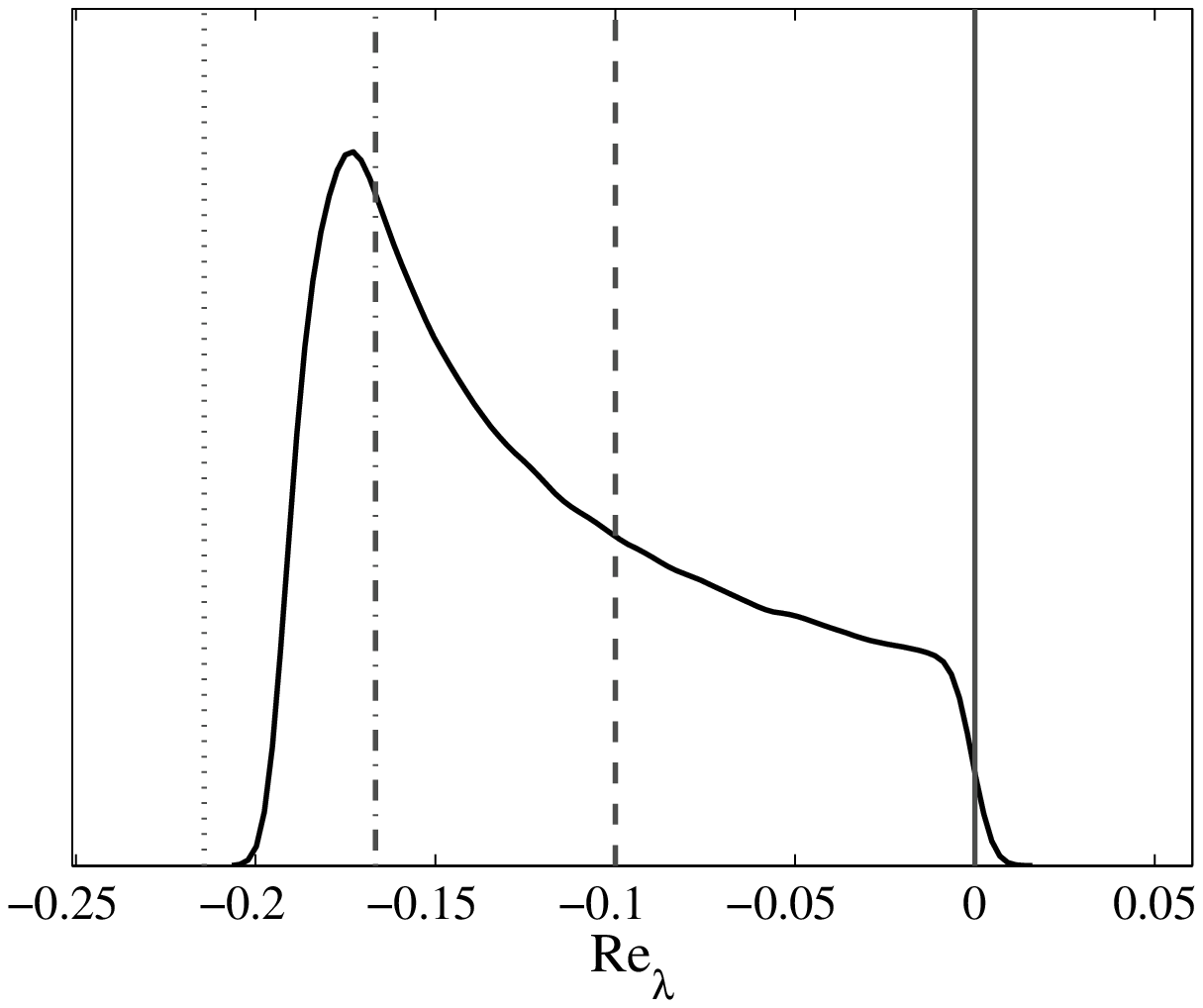}\\
 						\includegraphics[width=0.28\linewidth]{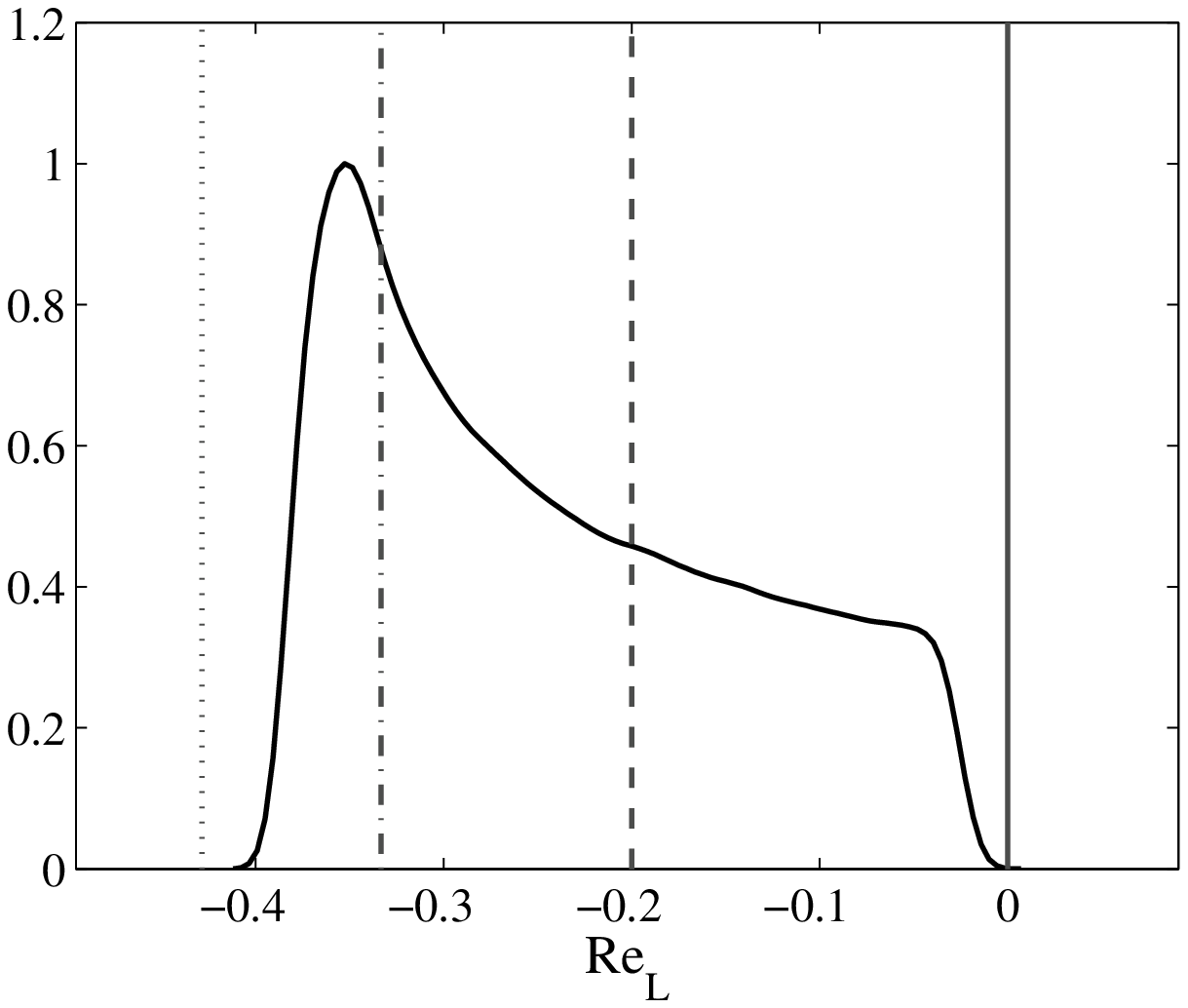} 
 						\includegraphics[width=0.28\linewidth]{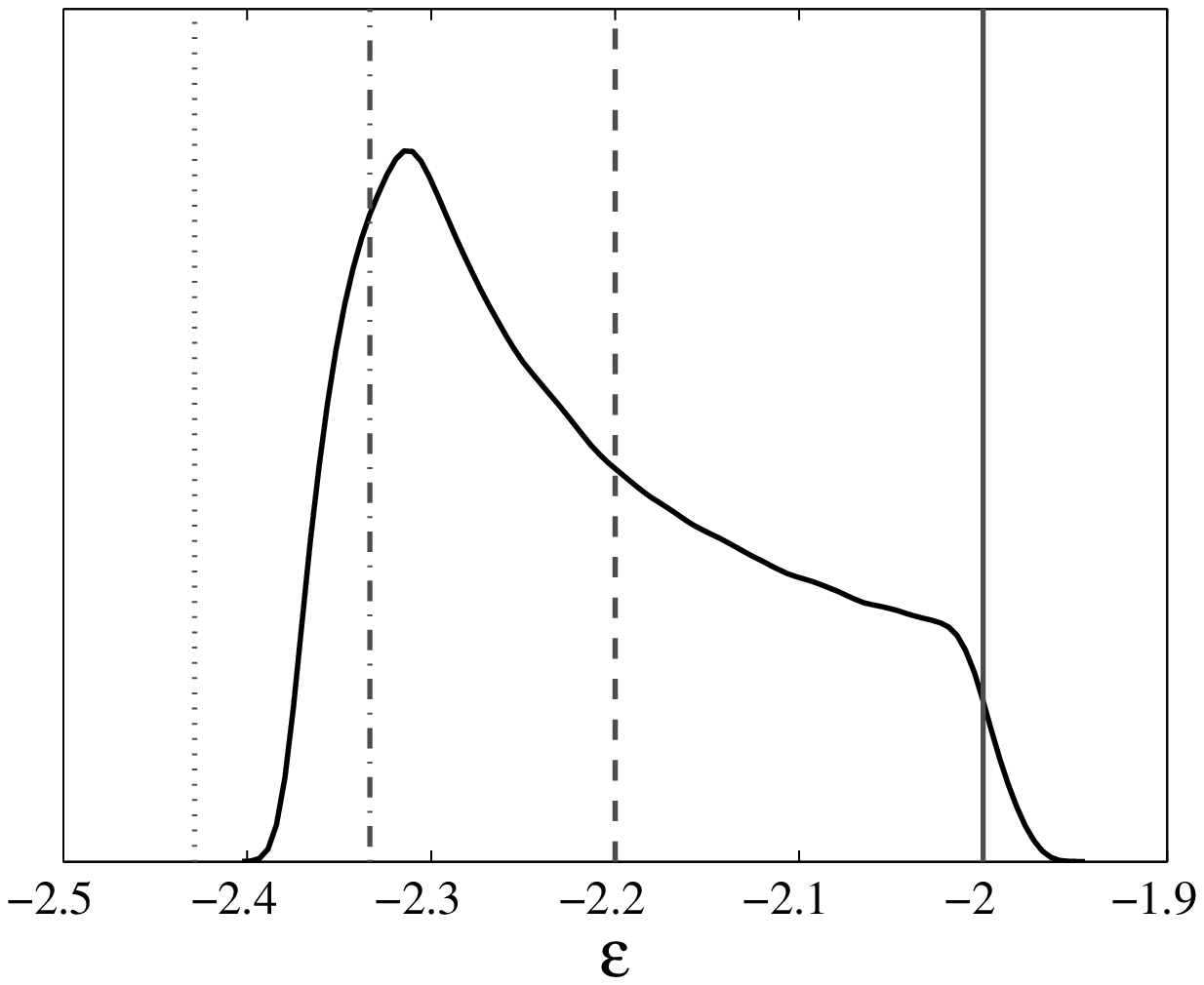}
 						\includegraphics[width=0.28\linewidth]{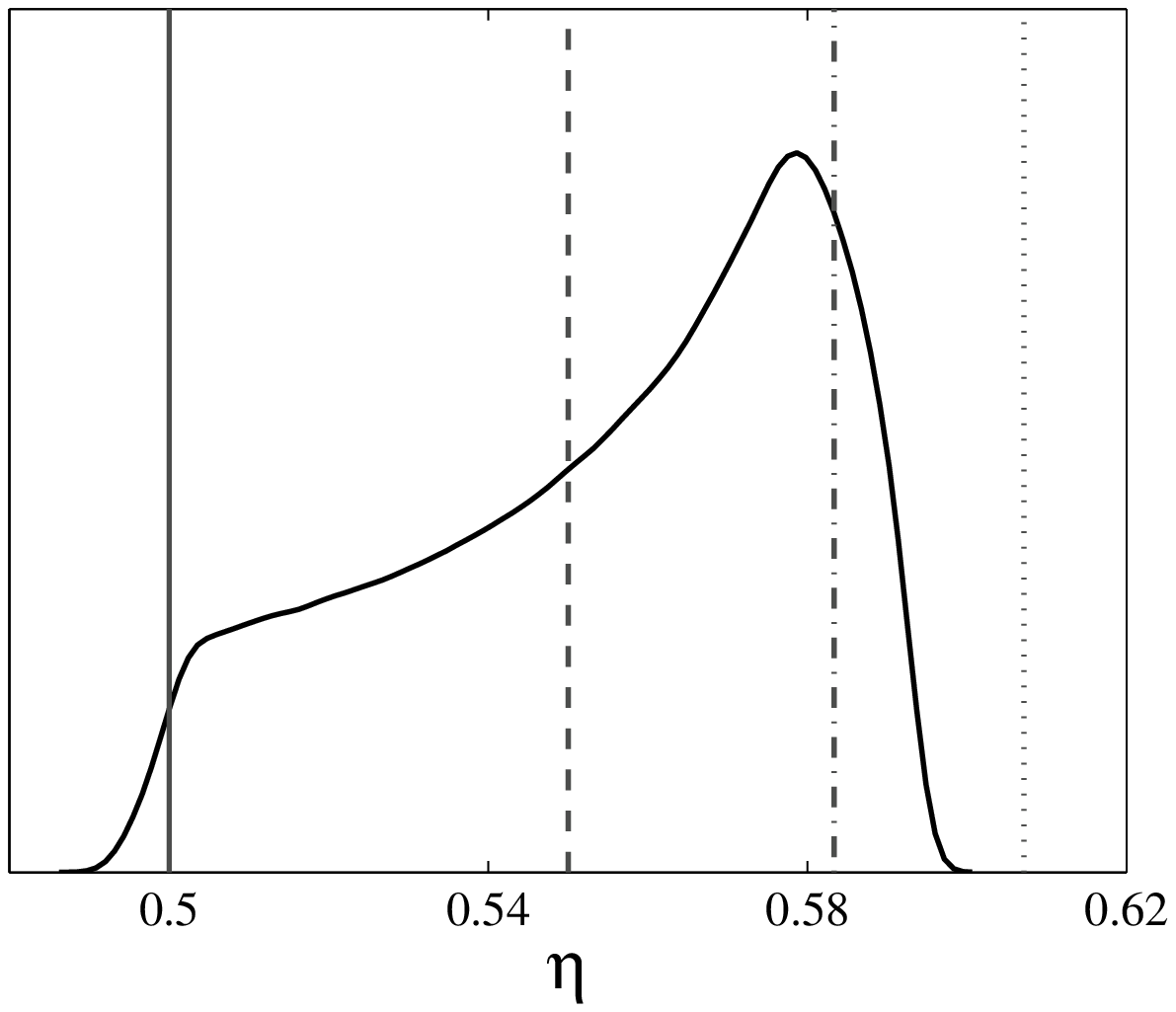}
 \caption{Pdfs of power-law exponents. Vertical lines refer to theoretical values retrieved from CBC analysis. Solid line: $\sigma = 1$; Dashed line: $\sigma = 2$ (Saffman turbulence); Dash-dotted line: $\sigma = 3$; Dotted line: $\sigma = 4$ (Batchelor turbulence).} 
 \label{fig::pdfCoeffHighRe10000} 
 \end{figure}

Uncertainty about the large-scale kinetic energy distribution at high Reynolds number is investigated first, considering $\sigma$, $p$ and $\alpha_5$ as random variables with {\em uniform} distribution.
Details are displayed in Table \ref{tab:uncVarHigh}. Let us emphasize that this case is typically representative of laboratory experiments in which the large-scale spectrum is neither controlled nor known.
In order to be sure that self-similar states are reached that corresponds to high-Reynolds number dynamics, the initial Reynolds number is set equal to $Re_\lambda (t=0) = 10^4$. With this value, a $Re_\lambda (t) > 400$ is obtained during the self-similar evolution of the system.
\begin{table}
\centering
\begin{tabular*}{\textwidth}{@{\extracolsep{\fill}} l c c c c }
\smallskip
Spectrum parameter                            & Mean value              &      Range                      & Reported variations                                  \cr  
\hline
$\sigma$				&   $2.5$		 	 &	$[1, 4]$			 & $[1, 4] \cup [+\infty]$		 	 \cr
$p$						&   $1.5$			 & 	$[1, 2]$		   & $[1.5, 2]$	 	 	\cr
$\alpha_5$			&   $4.1$		   &	$[3.9, 4.3]$ & $[3.967, 4.224]$		 	\cr
$\mu$				&   $0.25$		 	 &	$[0.2, 0.3]$			 & $[0.2, 0.3]$		 	 \cr
\hline
\end{tabular*}
\caption{Kinetic energy spectrum uncertain parameters, modeled as {\em uniform} random variables. Mean value and Range are related to present simulations, while Reported variations refer to variability reported in open literature.}
\label{tab:uncVarHigh}
\end{table}
Statistical results are displayed in Table \ref{tab::HighRe10000-1} and Fig. \ref{fig::pdfCoeffHighRe10000} which show parameters and probability density functions (pdf) of the power-law exponent  for  several global quantities, as computed from gPC/EDQNM-based response surfaces.
A very  significant   variability is observed on all exponents, which is mainly due to the large-scale exponent $\sigma$, as predicted by theoretical analysis. In the present case, the use of Sobol's coefficients (\cite{Sobol_1993}) shows that this parameter is responsible for $97-99\%$ of the total variance, indicating that at high Reynolds number $\sigma$ has a leading impact over the turbulence decay as already known by the seminal works of \cite{Batchelor1948}. The exponent $p$, which governs the smoothness of the energy spectrum peak, is observed to be the second leading parameter.
It is observed that the maximum variability (with $c_v$ close to 45\%)  is obtained for Reynolds number exponents, because of the combination of uncertainties on kinetic energy and the characteristic lengthscale under consideration, rendering the use of these parameters very complex in laboratory experiments.
Another important point is that most probable values $n_p$, which can be interpreted as the exponent values mostly to be observed in uncontrolled realizations of decaying turbulence, differ significantly from deterministic values, $n_d$, and the mean values computed from the response surface, $\bar n$. The two later values are seen to be quite close to each other, showing that $n_d$ may be substituted to $\bar n$  as a relevant surrogate  for faster analysis.\\
Comparisons with literature data are mostly restricted to the kinetic energy exponent, since other physical quantities have been much less documented.
A great variability is reported in experimental works: \cite{Benn1966} reported $-1.3 < n < -1.15$, while \cite{wl78} found $n \sim -1.34$. Hot wire measurents by \cite{Lavoie2007} are in the range $-1.22 \pm 0.02 < n < -1.04 \pm 0.02$. \cite{antonia2003} reported $n \sim -1.25$.
Lower values such as $n\sim -1$ can be found, e.g. \cite{Batchelor1948}.
The present range of variation $[n_{min}, n_{max}]$ is in very good agreement with values reported  by \cite{Moha1990} (e.g. Fig.13 in this reference). This is also true looking at the predicted most probable value $n_p \simeq -1.33$.
Lower values reported recently in \cite{Krog2010}, i.e. $n = -1,13 \pm 0.02$, are also encompassed by the present variation range.
The lowest predicted values is -1.384, indicating that lower value may be due to low-Reynolds number and/or finite domain size effects.
The same satisfactory agreement is found for other quantities. \cite{antonia2003} reported $n \sim -0.125$ for $Re_\lambda$, $n \sim 0.36-0.40$ for $L$.
Looking at Fig. \ref{fig::pdfCoeffHighRe10000} one observes that, for all physical quantities, the pdf does not encompass the theoretical value associated with Batchelor turbulence ($\sigma =4$).
This is also coherent with results displayed in Table \ref{tab::HighRe10000-1}. In fact, this is related to the existence of a time drift of the prefactor in that case, and to previous observations reported by many authors, who observed a violation of the hypothesis of permanence of large eddies. 
Exponent value associated with $\sigma =1$ have a very low probability, indicating that this case is very difficult to identify using real-life experiments. Interestingly, most probable exponent values are close to those of $\sigma =3$.
An important conclusion here is that it is very difficult to derive the distribution kinetic energy at large scales using power-law exponents of global quantities, and that direct measurements of the spectrum is the only safe way to get conclusive results on that point.
\begin{table} \centering \begin{tabular*} {\textwidth}{@{\extracolsep{\fill}} l c c c c c c c c} \smallskip
Exponent &$n_{d}$& $\bar{n}$& $c_v$& [$n_{min}$,$n_{max}$] & $n_{p}$ & $\varsigma_{\sigma}$ & $\varsigma_{p}$ & $\varsigma_{\alpha_5}$\cr \hline
$q$            &-1.242   &-1.227  & $8.6\%$    &[-1.384, -0.980]   &-1.330 & 9.85e-001   &1.40e-002  &3.73e-006 \cr
$L$            &0.363  &0.374  &$13.4\%$         &[0.304, 0.475]   &0.314 &9.99e-001  &9.46e-004  &1.93e-006 \cr
$Re_{\lambda}$ &-0.128  &-0.118 & $45.6\%$  &[-0.194, 0]  &-0.173 &9.92e-001  &7.05e-003 &1.18e-006 \cr
$Re_{L}$       &-0.258 &-0.240& $42.9\%$ &[-0.388, -0.017] &-0.353 &9.94e-001  &5.69e-003 &1.63e-007 \cr
$\varepsilon$  &-2.227  &-2.218 & $4.7\%$  &[-2.380, -1.968]  &-2.315 &9.76e-001  &2.36e-002 &8.09e-006 \cr
$\eta$         &0.557  &0.554 &   $4.7\%$       &[0.492, 0.595]  &0.579 &9.76e-001  &2.36e-002 &8.09e-006 \cr \hline
 \end{tabular*} \caption{Statistical data  related to power-law exponents.  $n_{d}$, $\bar{n}$,  $c_v$, $n_{min}$, $n_{max}$, $n_{p}$ and $\varsigma$ refer to the: deterministic value associated with the mean value of the uncertain parameters, statistical mean value, coefficient of variation (standard deviation referred to $\bar n$), minimum value, maximum value,  most probable value (i.e. peak of the pdf) and the normalized partial variance associated to the subscript variable, respectively.} \label{tab::HighRe10000-1} \end{table}
\section{Saffman and Batchelor turbulence stochastic analysis}
\label{bottleHigh}
\begin{figure} \centering \includegraphics[width=0.275\linewidth]{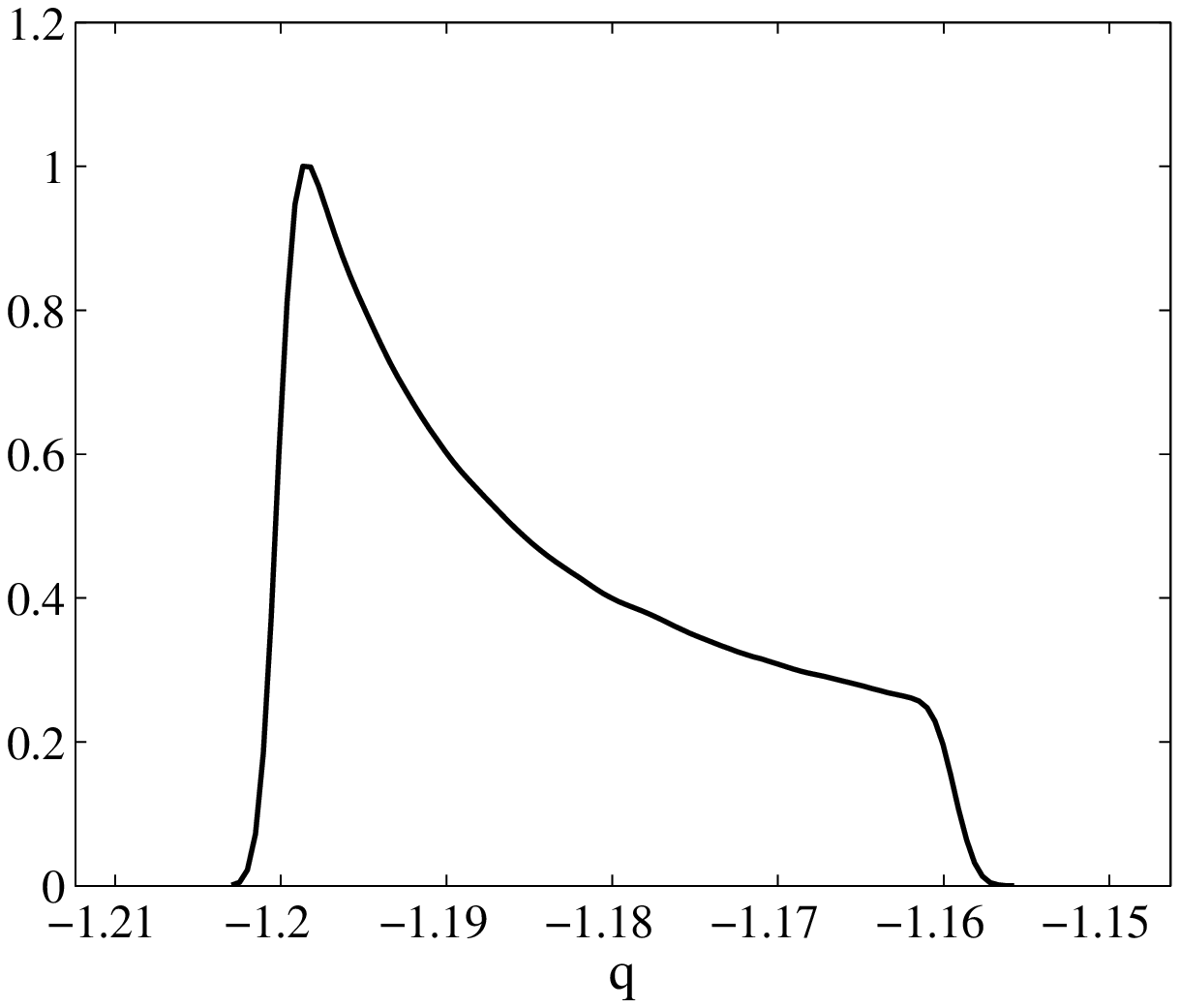} 
 						\includegraphics[width=0.275\linewidth]{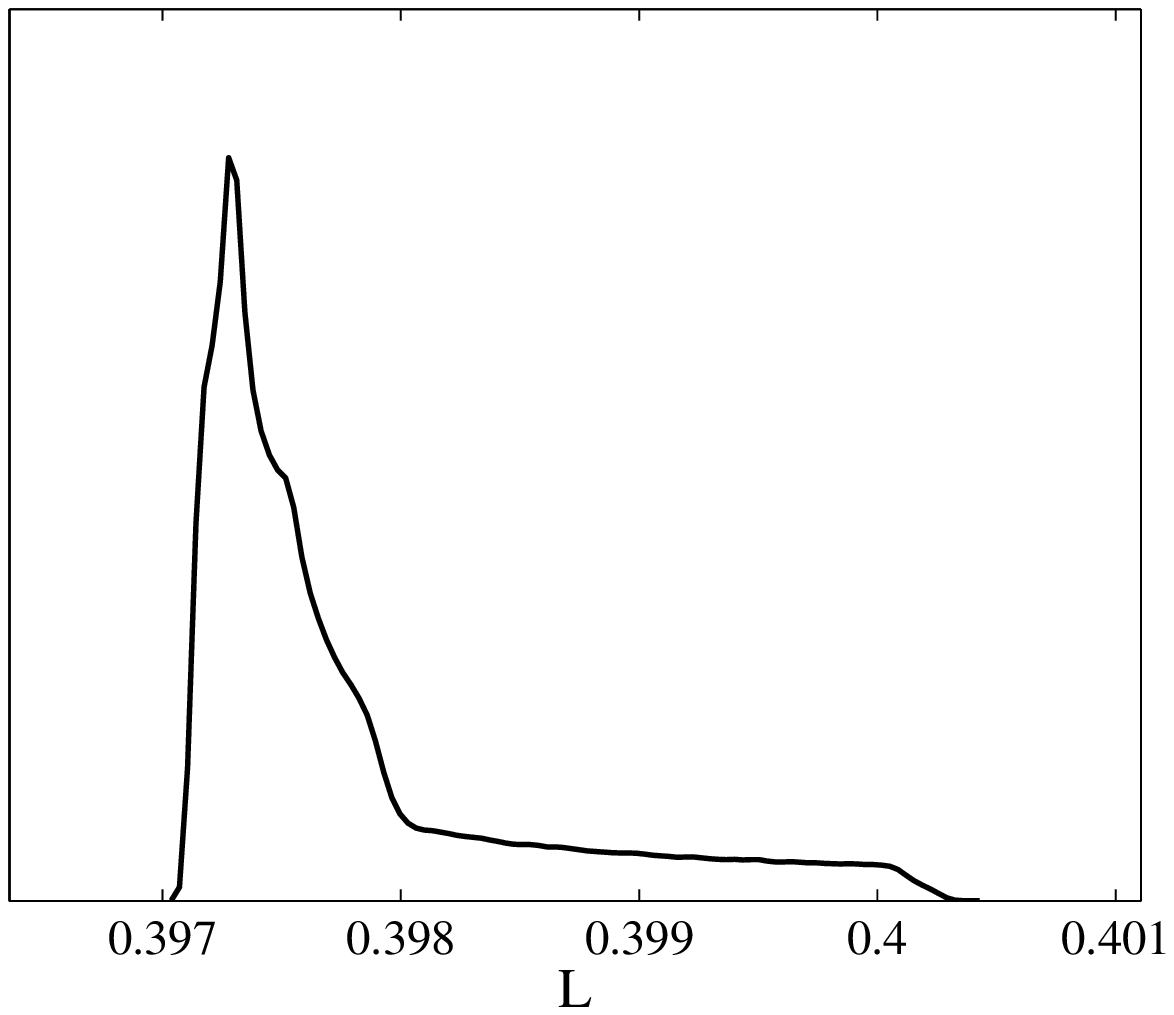}
 						\includegraphics[width=0.275\linewidth]{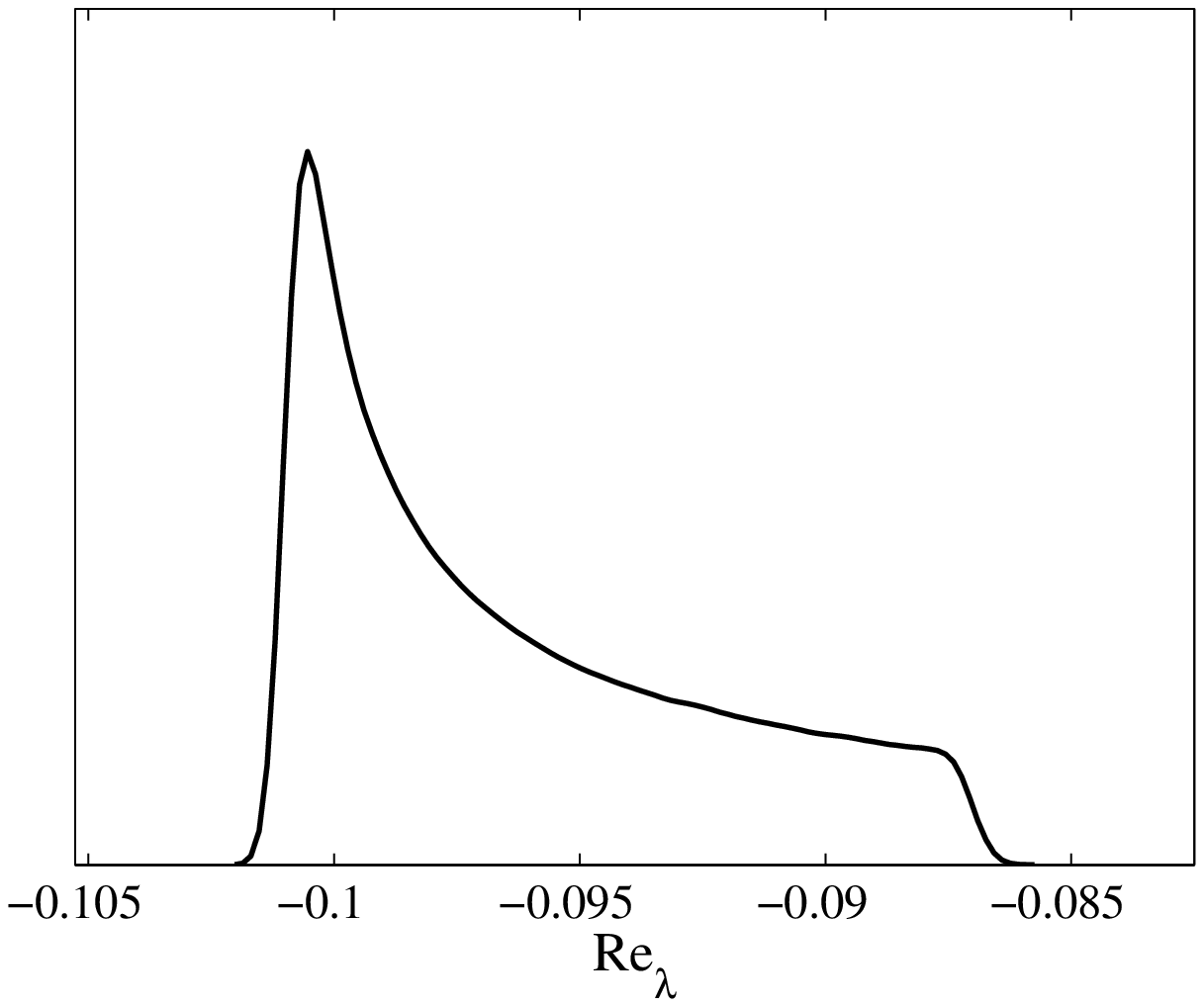}\\
 						\includegraphics[width=0.275\linewidth]{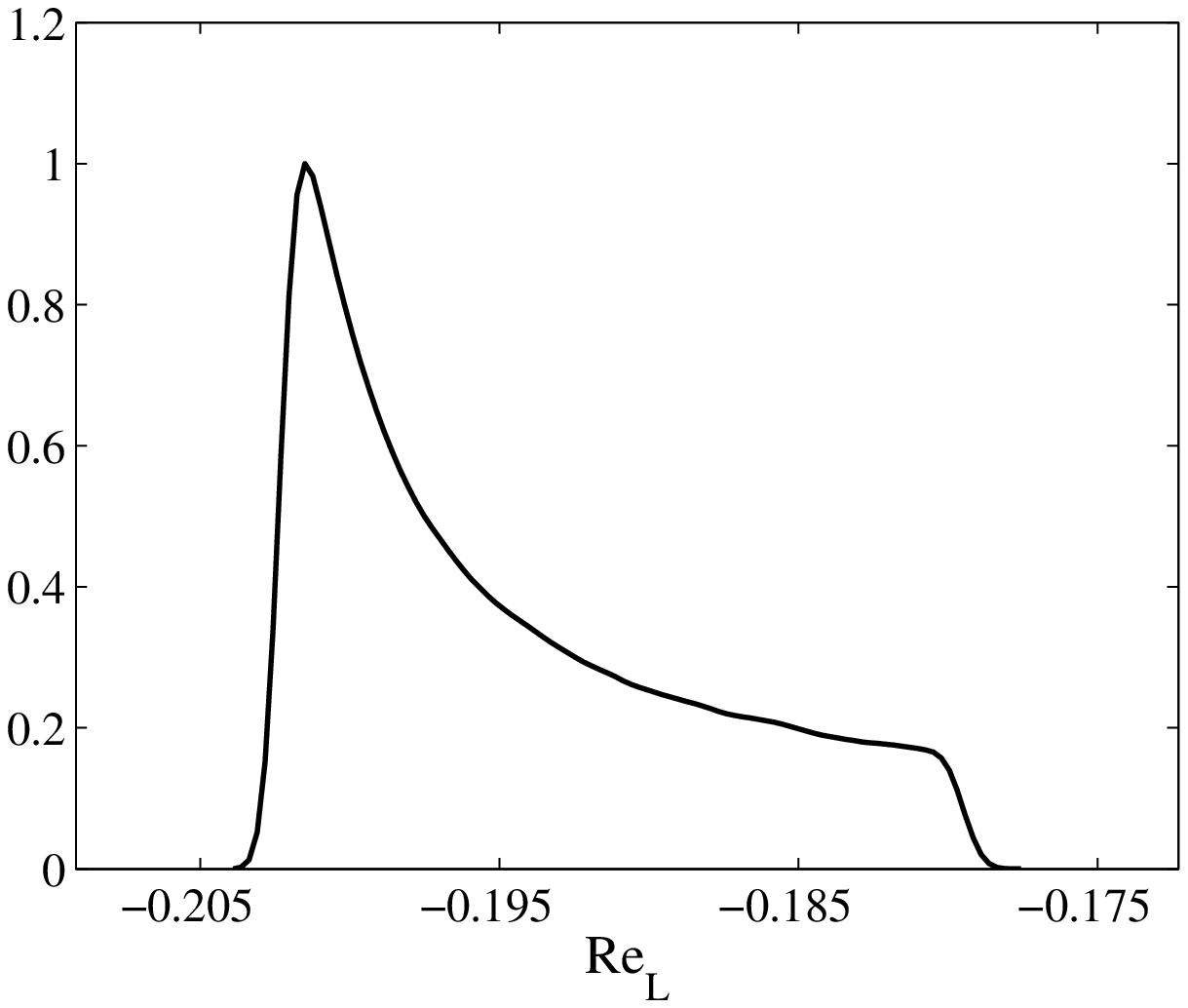} 
 						\includegraphics[width=0.275\linewidth]{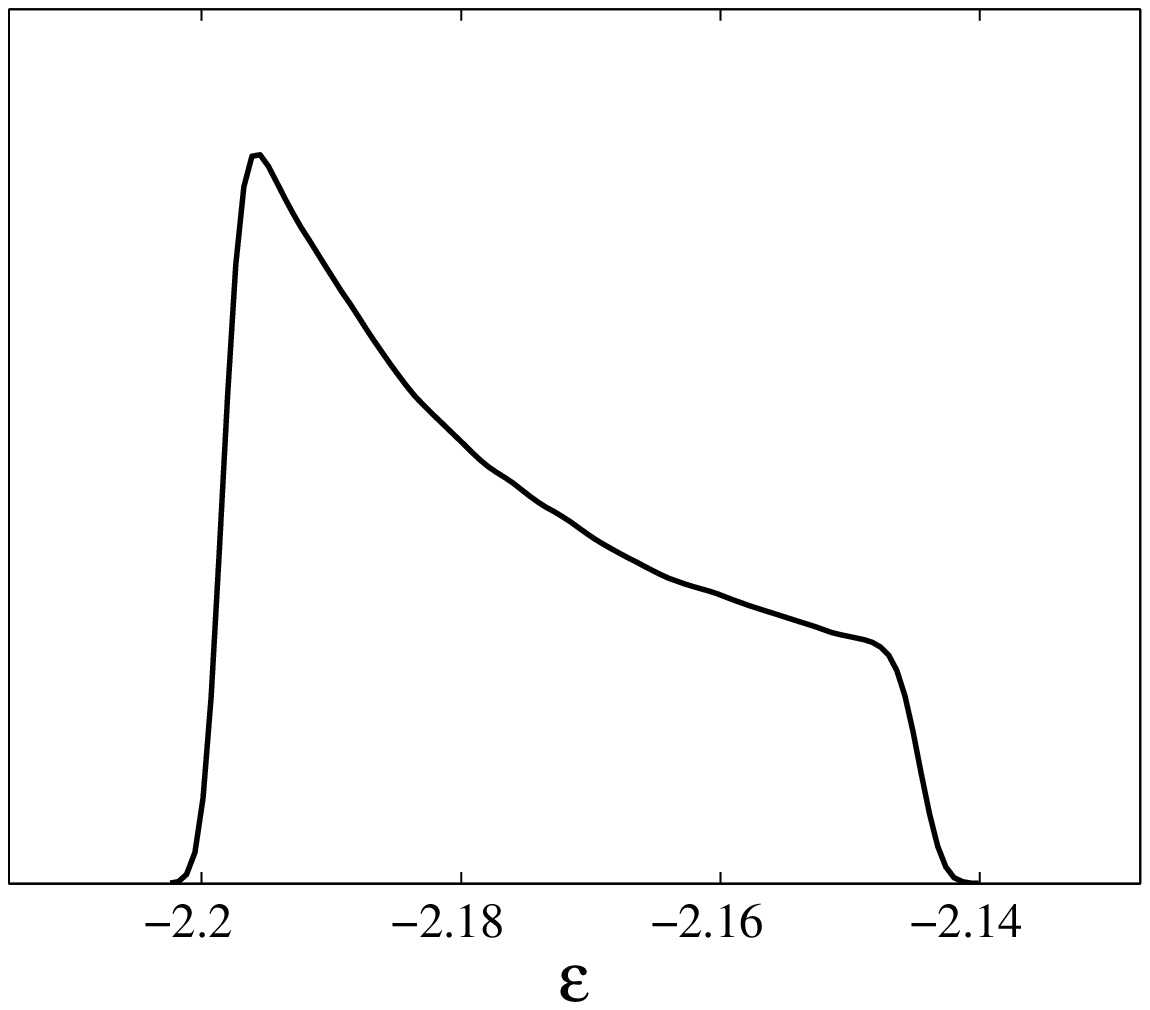}
 						\includegraphics[width=0.275\linewidth]{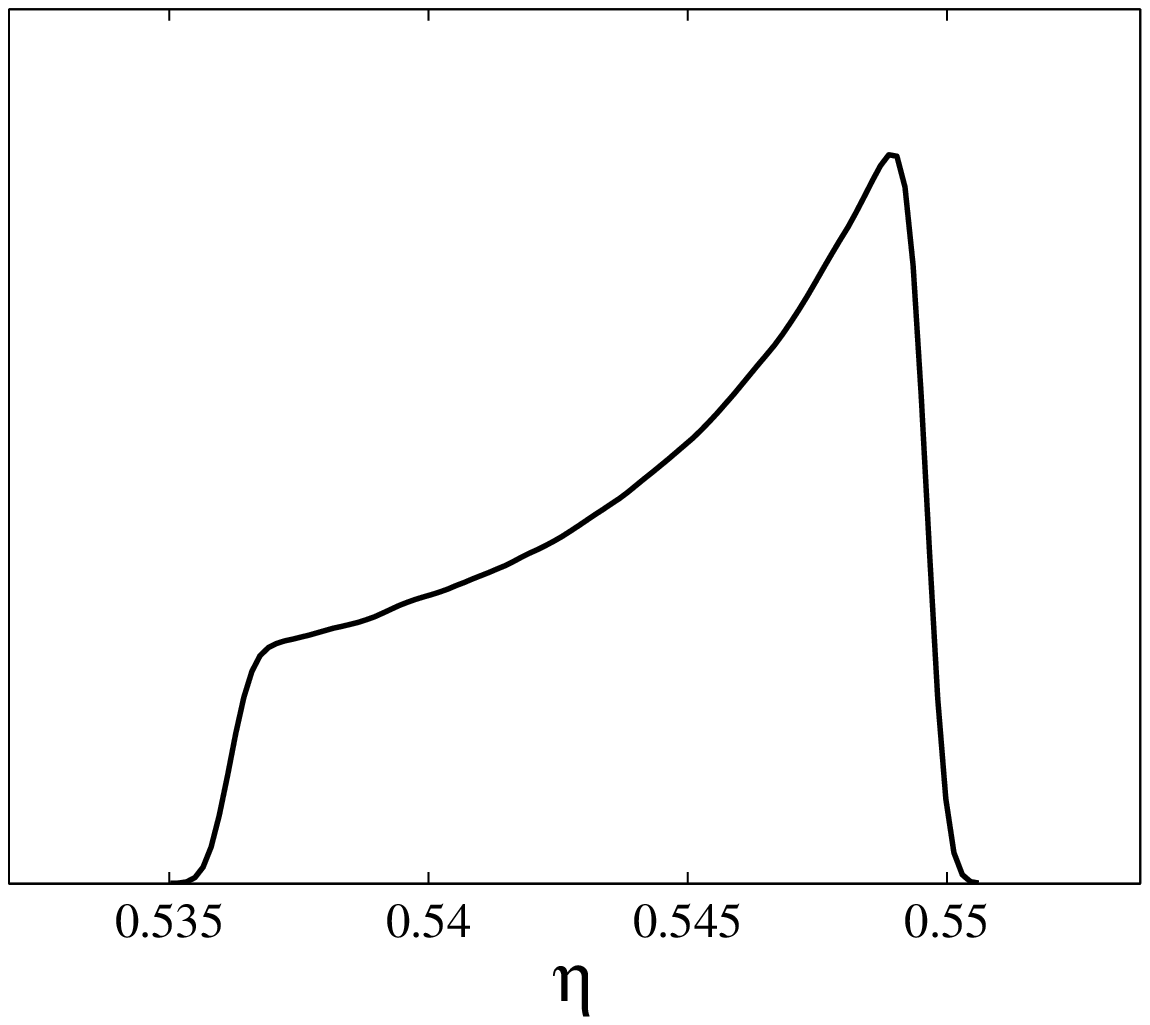} \caption{Power-law exponents pdf for Saffman turbulence ($\sigma = 2$) for $ Re_{\lambda} (t) > 400$.} \label{fig::figure2} \end{figure}

\begin{figure} \centering \includegraphics[width=0.275\linewidth]{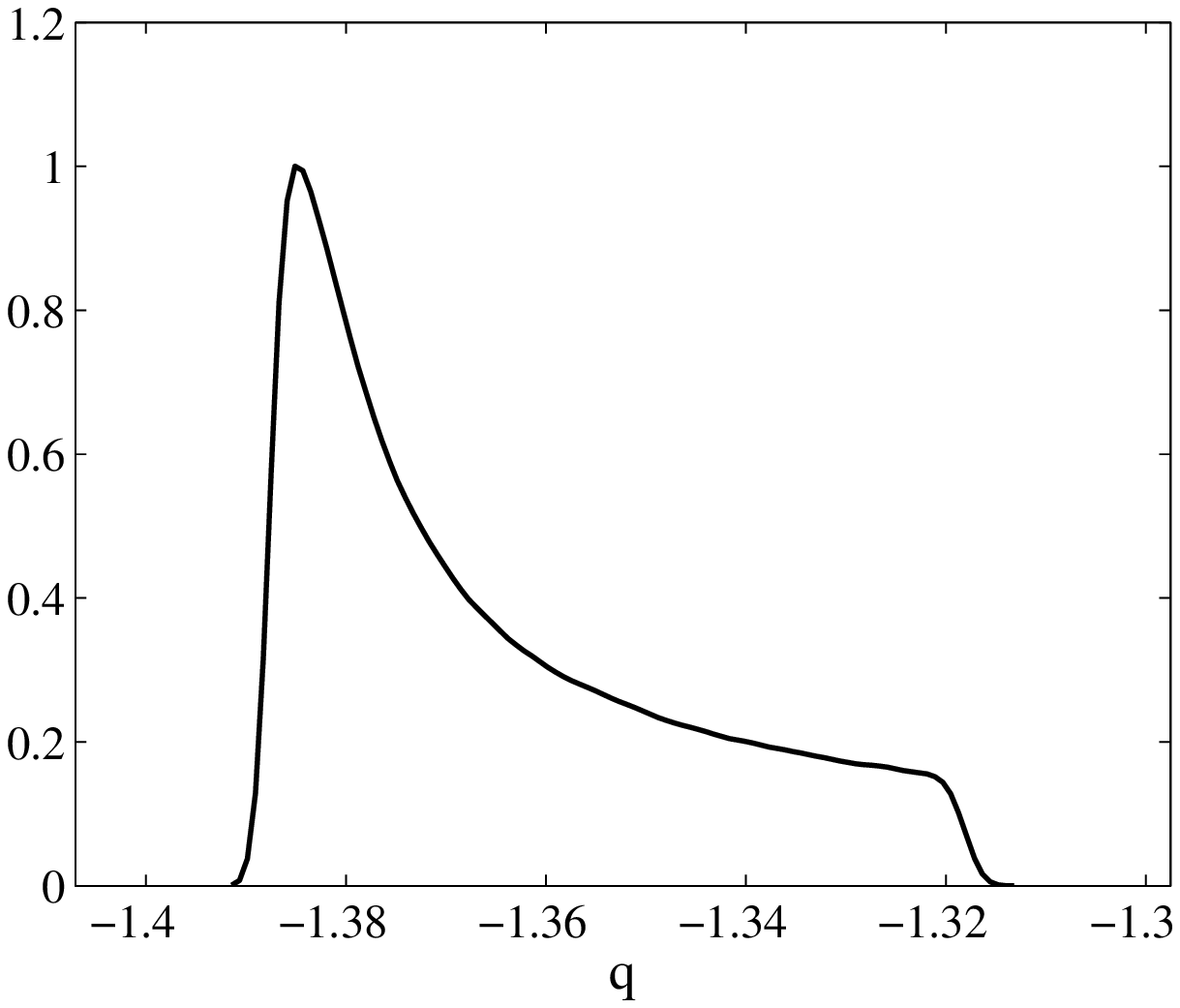} 
 						\includegraphics[width=0.275\linewidth]{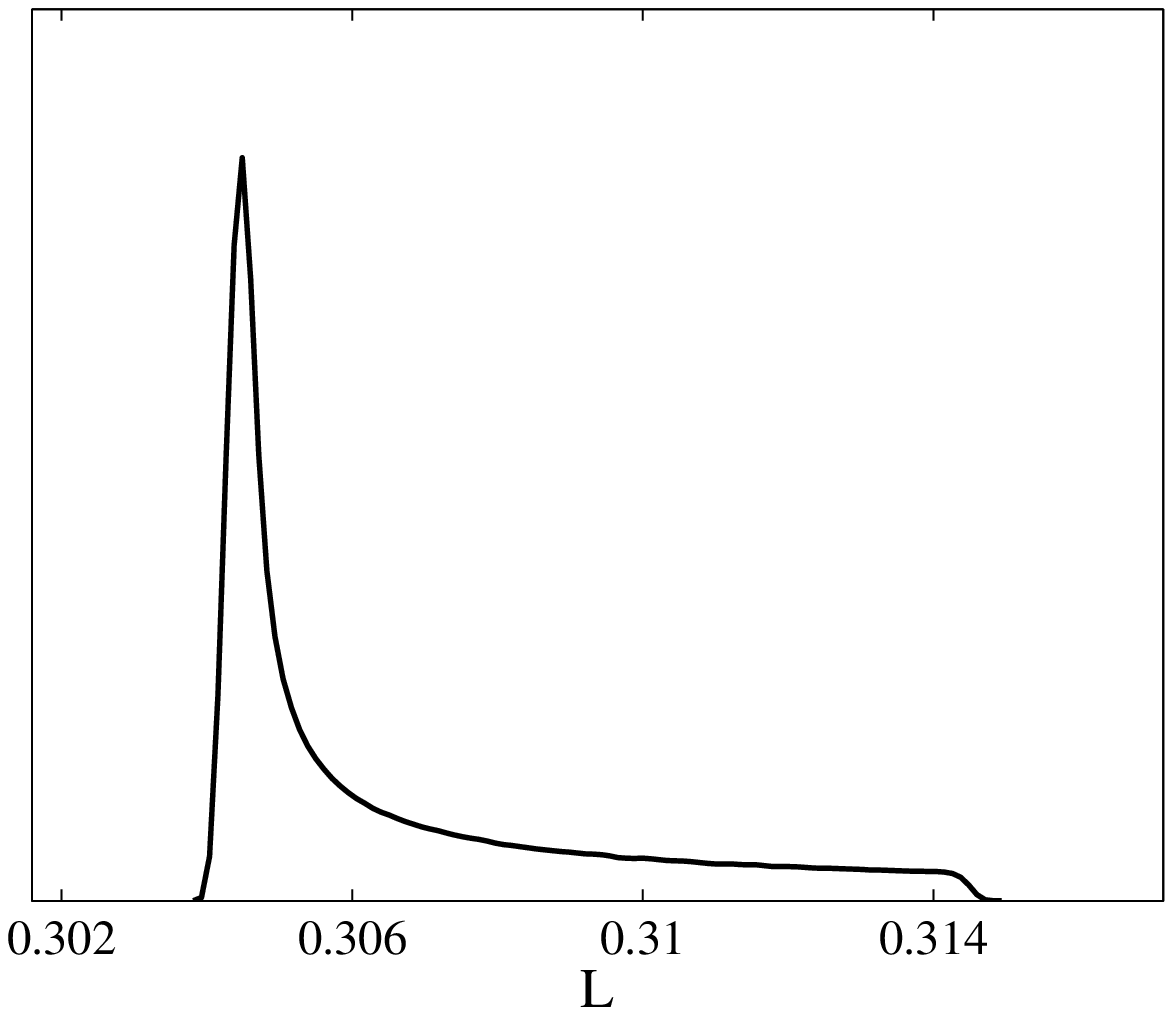}
 						\includegraphics[width=0.275\linewidth]{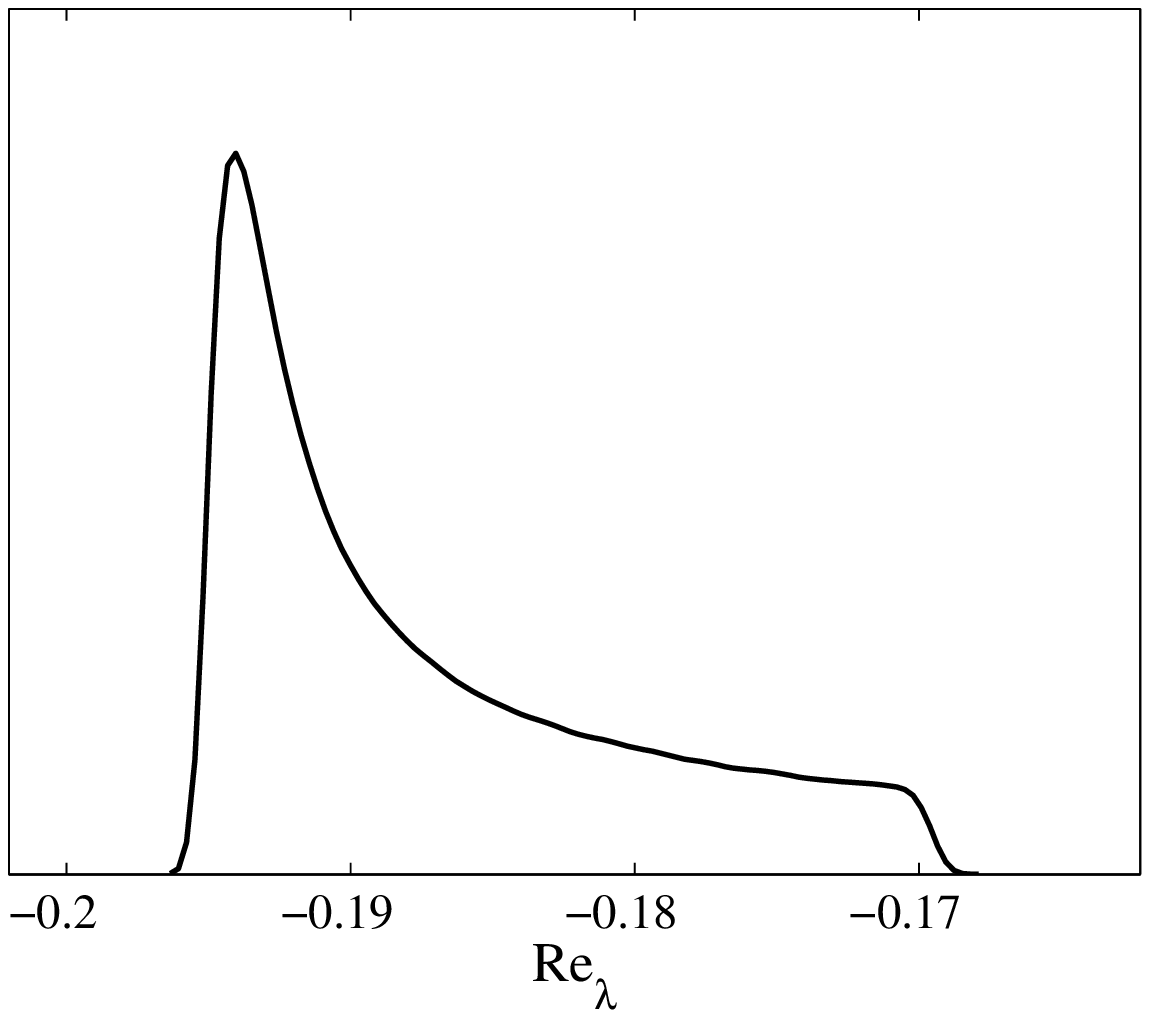}\\
 						\includegraphics[width=0.275\linewidth]{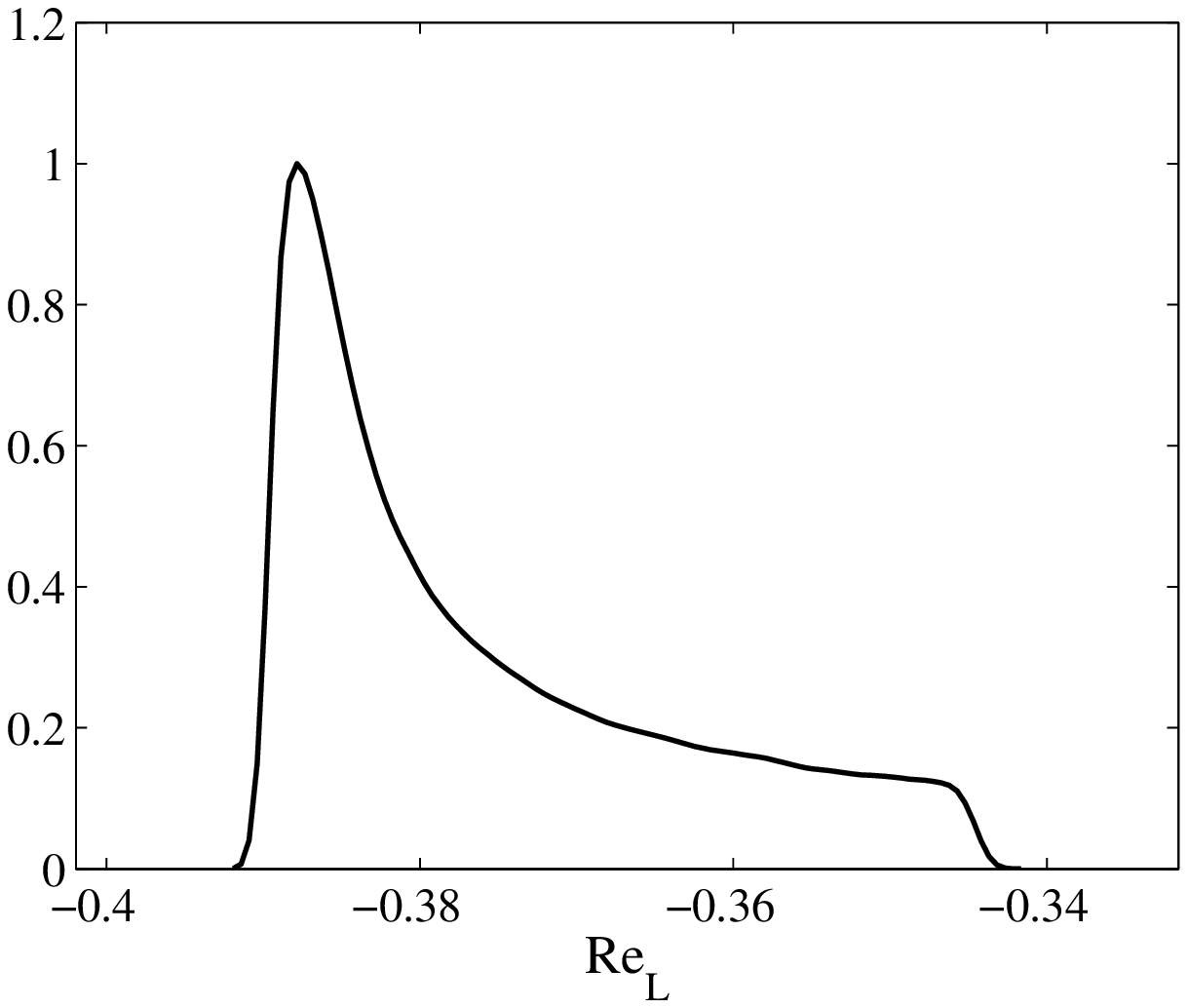} 
 						\includegraphics[width=0.275\linewidth]{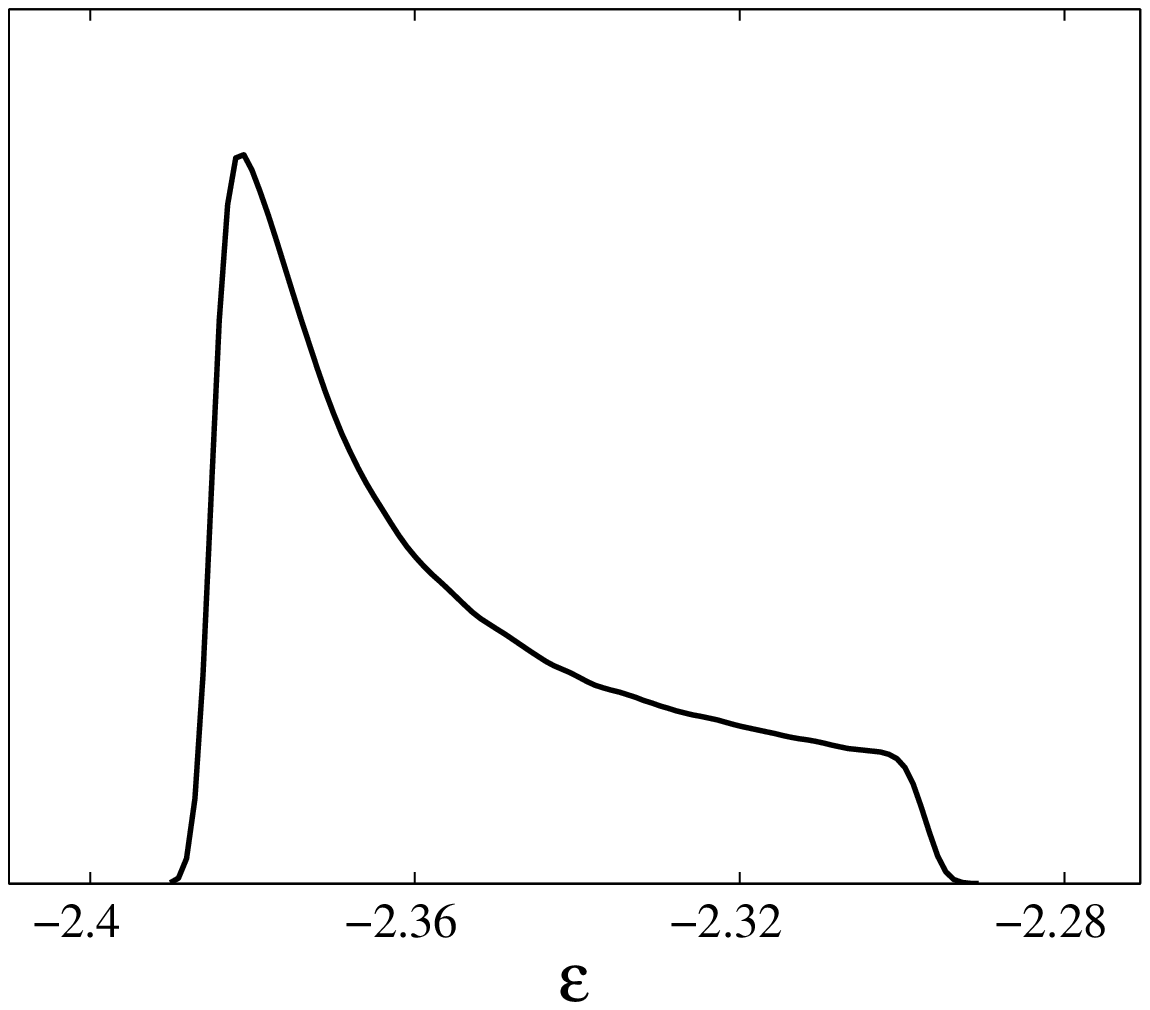}
 						\includegraphics[width=0.275\linewidth]{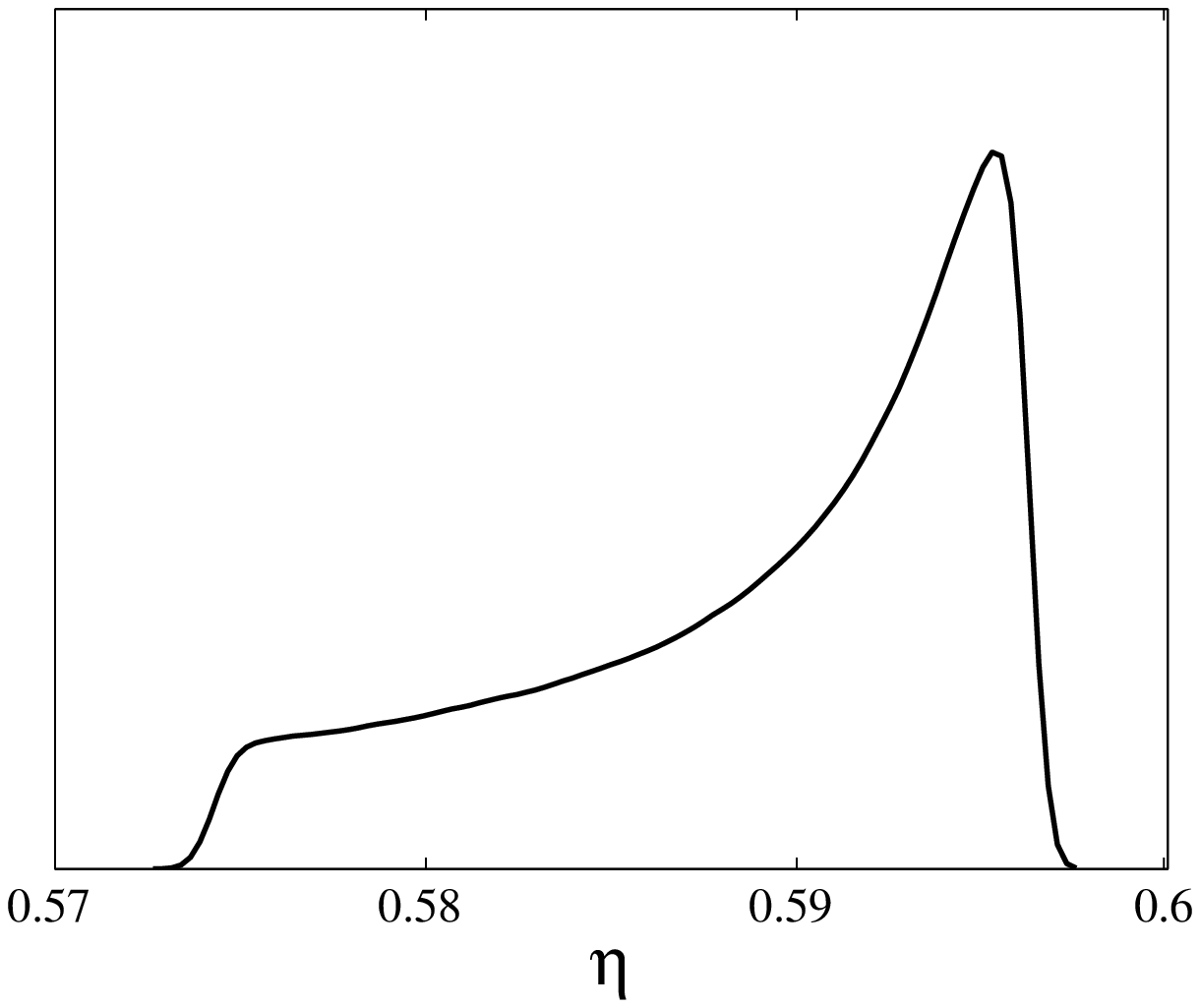} \caption{Power-law exponents pdf for Batchelor turbulence ($\sigma = 4$) for $Re_{\lambda} (t) > 400$.} \label{fig::figure3} \end{figure}

We now consider uncertainty for fixed deterministic $\sigma$ values, analysing the two canonical cases known as Saffman turbulence ($\sigma = 2$) and Batchelor turbulence ($\sigma =4$). The stochastic approach is developed considering the set of parameters $p$, $\alpha_5$ and $\mu$ as \textit{uniform} random variables (see Table \ref{tab:uncVarHigh}). The results comply with the correspective conservation laws and in the case of Saffman turbulence the theoretical relation correlating $n_{\varepsilon}$ and $n_{L}$ is recovered with an error lower than $2\%$.
The stochastic analysis of low-order moments (e.g. coefficient of variation $c_v$) indicates that small variance are recovered for both cases, up to $4\%$ for the Reynolds number. Sensitivity analysis indicates that initial conditions depends almost exclusively on $p$. However, extreme events (e.g. maximum and minimum power law exponents) are instead quite distant one from the other, spanning from values close to the theoretical ones to values even $10-15\%$ lower in magnitude. The same trend is noticeable in the pdf reported in Fig. \ref{fig::figure2} and \ref{fig::figure3}, where the most probable value is always quite close to the theoretical one: the introduction of uncertainties over the model constant of the energy spectrum functional form leads to a high probability of recovering a turbulence decay state whose characteristics are close to the theoretical value, but the probability of a significant gap is not negligible. In particular when the width of the inertial range is reduced (wide and smooth spectrum peak for small values of $p$), the turbulence decay is characterised by a lower magnitude of the power law coefficients. On the contrary at  high $p$ values, i.e. if the energy spectrum shape is similar to CBC spectrum model,  power law exponents get closer to the theoretical ones.
An important remark is that ranges of variation of $n$ do not overlap. Therefore, tentative values of $\sigma$ can be deduced from $n$ in high Reynolds number, unbounded isotropic turbulence inverting CBC formula (see Table \ref{tab::CBC-1}). 
$n \sim -1.3$ should be associated with Batchelor turbulence, while $-1.2 \leq n \leq -1.15$ may indicate the occurrence of Saffman-type turbulence.
%
%\begin{figure} \centering
% 						\includegraphics[width=0.7\linewidth]{./figures/batch_spec.eps}
% 						 \caption{Initial energy spectrum sensitivity to $P$ in the case of Saffman turbulence.} \label{fig::spec} \end{figure}
						 
%
\begin{table} \centering \begin{tabular*} {\textwidth}{@{\extracolsep{\fill}} l c c c c c c} \smallskip
 &$n_{d}$& $\bar{n}$& $c_v$& $[ n_{min}, n_{max}]$ & $Q_{95}$ & $ n_{p}$ \cr 
 \hline
&  \multicolumn{5}{c}{Saffman turbulence ($\sigma=2$)} \cr \hline
$q$            & -1.188   &-1.185   &1\% &[-1.200,   -1.158] &-1.163 &-1.199 \cr
$L$            & 0.397    &0.398    &0.2\% &[0.397,   0.400]   &0.400  &0.397 \cr
$Re_{\lambda}$ & -0.097   &-0.096   &4.2\% &[-0.101,  -0.087]  &-0.088 &-0.101 \cr
$Re_{L}$       & -0.196   &-0.195   &3.3\%  &[-0.203, -0.179]   &-0.182 &-0.202 \cr
$\varepsilon$  & -2.181   &-2.177   &0.7\%  &[-2.199,  -2.144]  &-2.149 &-2.195 \cr
$\eta$         & 0.545    &0.544    &0.7\%  &[0.536,  0.550]    &0.549  &0.549 \cr
 \hline &  \multicolumn{5}{c}{Batchelor turbulence ($\sigma = 4$)} \cr \hline
$q$            & -1.370   &-1.365   &1.4\% &[-1.388, -1.317]   &-1.326 &-1.385 \cr
$L$            & 0.305    &0.307    &0.9\% &[0.304,  0.315]    &0.313  &0.304 \cr
$Re_{\lambda}$ & -0.189   &-0.187   &3.8\% &[-0.195, -0.169]   &-0.173 &-0.194 \cr
$Re_{L}$       & -0.380   &-0.376   &3.3\%  &[-0.390, -0.344]   &-0.350 &-0.388 \cr
$\varepsilon$  & -2.362   &-2.355   &1\% &[-2.385, -2.295]   &-2.307 &-2.381 \cr
$\eta$         & 0.590    &0.589    &1\% &[0.574, 0.596]     &0.596  &0.595 \cr \hline
\end{tabular*} \caption{Statistical data  related to power-law exponents, in case of large-scale uncertainty at fixed values of $\sigma$. Captions: see Table \ref{tab::HighRe10000-1}.} \label{tab::bottleSigma2Re10000-1} \end{table}
\section{Concluding remarks }
The present results, based on EDQNM closure for the Lin equation and not on the sole dimensional analysis,  confirm that the slope of the spectrum at large scales at initial time is the leading parameter that governs the decay regime. 
A $q \sim t^{-1}$ decay can be observed at finite time and finite Reynolds number for $\sigma \simeq 1$, a small variability with other large-scale parameters being observed. In that regime, turbulence decays at almost constant $Re_\lambda$. The occurrence of this regime shows that $n=-1$ is not tied to the existence of a universal asymptotic regime, and that experimental evidence of the existence of such a universal regime may be very difficult to obtain, unless $\sigma$ is accurately and independently measured.
The dependency upon $\sigma$ is observed to remain, even at Reynolds numbers considered in the present study ($Re_\lambda > 400$), which are higher than those considered in almost all existing published results. The variability observed for Batchelor and Saffman turbulence with respect to details of spectrum shape is relatively small, leading to an almost univoque identification of $\sigma$ from $n$, inverting relations retrieved from the CBC theory, at least at high Reynolds number and restricting the analysis to integer values for $\sigma$.
Present results show no evidence of the existence of a universal regime with $q \sim t^{-1}$ at high but {\em finite} Reynolds number, even looking at long-time evolution.
Another point is that, even though isotropy is perfectly satisfied and finite size effects are absent (the ratio integral lengthscale / lowest resolved scale is greater than 250), the decay exponent is governed by $\sigma$. Therefore, convergence toward a single value reported in some experimental works when isotropy is refined (e.g. \cite{Lavoie2007}) might be the signature of the large-scale spectrum produced by a given experimental set-up rather than a true universal value.

\end{document}